# Polarization controlled second harmonic generation imaging of stretched collagen fibrils reveals collagen's deformation pathway in situ

MacAulay Harvey[1], Konstantin Röder[2], Richard Cisek[1], Danielle Tokarz[1,†] , Laurent Kreplak[3,*]

[1]Department of Chemistry, Saint Mary's University, 923 Robie Street, Halifax, NS, B3H 3C3, Canada

[2]Randall Centre for Cell & Molecular Biophysics, King's College London, Great Maze Pond, SE1 9RL, London, United Kingdom

[3]Department of Physics and Atmospheric Science and School of Biomedical Engineering, Dalhousie University, Halifax, NS, B3H 4J5, Canada

[†]danielle.tokarz@smu.ca

[*]kreplak@dal.ca

## Abstract

The tensile properties of single collagen fibrils, the building block of load-bearing tissues, have been studied extensively by nanomechanical techniques and molecular dynamics simulation. However, the deformation pathway of collagen molecules within fibrils has not yet been observed experimentally. In addition, the role played by divalent and trivalent enzymatic crosslinks in modulating this deformation pathway is poorly understood. Here we used polarization controlled second harmonic generation (SHG) imaging combined with atomic force microscopy (AFM) to characterize the molecular state of collagen triple helices within stretched single collagen fibrils. The fibrils were extracted from a pair of bovine leg tendons from the same animal in order to compare fibrils with a high amount of immature divalent crosslinks to fibrils with a high amount of mature trivalent crosslinks. By selecting fibrils with a large SHG intensity gradient along their length and then imaging the same fibrils by AFM we were able to link the observed intensity gradient with a gradient in D-band strain and a gradient in molecular strain as estimated from the SHG anisotropy parameter $\rho$. In contrast to previous studies at the tendon scale, we observed that the SHG molecular strain is always larger than the D-band strain for all fibrils with this difference being largest for fibrils rich in divalent crosslinks. By analyzing the behavior of the relative density of SHG emitters as a function of molecular strain, we observe a two-state transition from an SHG producing to a non SHG producing state with a free energy barrier between 6 and 10 $k_BT$ that we propose corresponds to the local untwisting of the collagen triple helix superhelical twist which likely preceded bond rupture and loss of the SHG signal. We also show that trivalent crosslinks tend to delay the transition onset compared to divalent crosslinks.

## Introduction

Collagen fibrils are nanoscale biological ropes which serve as the primary load bearing elements of the extracellular matrix (ECM). The basic building block of the collagen fibril is the collagen molecule, which is a triple helix ~300 nm in length and ~1 nm in width[1,2]. Within fibrils, molecules are assembled in staggered arrays, leading to a characteristic density fluctuation along the fibril, known as the D-band, with a periodicity of ~67 nm[3]. Because of their important role in determining the large-scale mechanical properties of tissues there has been a significant body of previous work on the mechanical properties of individual collagen fibrils in indentation and tension that was recently reviewed here[4]. Collagen fibril mechanics is of particular importance since mechanical interactions between cells and collagen have been shown to play several important roles in cell signaling within the ECM[5–10]. Therefore, increasing our understanding of collagen mechanics helps to increase our understanding of the overall architecture of the ECM, and will be of high importance in the development of collagen-based scaffolds for tissue engineering applications.

The tensile response of collagen fibrils has been directly probed in a few different ways using atomic force microscopy (AFM), microelectromechanical devices and homebuilt nanotensile testers[11]. These tests have revealed three distinct phases of the collagen mechanical response with phase I from 0 – 7 % strain, phase II from 7 – 15 % and phase III from 15 % until failure[12]. The deformation mechanisms associated with each of these phases have been investigated using coarse grained and atomistic molecular dynamics. These computational studies have revealed that within the first phase the dominant deformation mechanism is unkinking and uncoiling of the triple helices and stretching of the molecular backbone[13,14]. The same deformation mechanism was proposed independently by Misof et al. when analyzing the tensile properties of a rat tail tendon while performing X-ray scattering experiments[15]. Within phase II the strain on the fibril is able to overcome intermolecular interactions, and molecular sliding is expected to be the dominant deformation mechanism. For fibrils lacking crosslinks the molecules will eventually slide apart in phase II resulting in rupture of the fibril, however the presence of intermolecular crosslinks can in principle restrict this sliding leading to phase III in which the dominant deformation mechanism is stretching of the crosslinks. The number and type of intermolecular crosslinks is an important factor in determining the mechanical behavior of a fibril with trivalent (mature) crosslinks being better able to restrict molecular sliding than divalent (immature) crosslinks. This concept was recently recapitulated in a finite element model of a single collagen fibril incorporating different types of crosslinks[16]. From an experimental standpoint the best models to study the respective roles of divalent and trivalent crosslinks are the leg tendons of equines and ovines[17]. In these systems the extensor tendons have a positional role and are rich in divalent crosslinks whereas flexor tendons have an energy storing role and are rich in trivalent crosslinks [17–19].

While computational modeling has provided some very valuable insights into the mechanical properties of collagen fibrils (e.g.[14,20]), there is currently a lack of direct experimental evidence for these deformation mechanisms. This is mainly due to the limitations of available analytical techniques for studying collagen fibrils under tension. Mechanical testing provides direct measurements of the mechanical properties of the fibril over a large range of strain values; however, it does not provide any direct structural information. AFM imaging of fibrils after a stretch and release cycle[21], or which have been held under tension on an elastic substrate[22,23] allows high resolution measurements of the radial modulus, as well as measurements of the D-band extension. However, while measurements of the D-band strain have been used as a proxy for the strain on the fibril[15,23–25], this likely underestimates the true strain applied to the fibril[26]. Additionally, AFM measurements provide no direct information on the molecular structure of the collagen triple helices. Wide-angle X-ray scattering has been used to directly measure the molecular strain in collagen; however, this technique is limited to relatively low strain values of < 2.5 %[27,28]. Therefore, an experimental technique with the ability to measure molecular strain in collagen over a wider range of strain values would be highly valuable in investigating the mechanical properties of collagen fibrils.

Polarization-resolved second harmonic generation microscopy (PSHG) is a nonlinear optical microscopy technique which allows label free visualization of noncentrosymmetric structures such as the collagen triple helix. This technique has been utilized extensively for investigations of the organization of collagen within tissue samples, mainly towards applications in automated pathology[29–31]. PSHG allows accurate measurements of the pitch angle of helical molecules[32], and has been shown to obtain pitch angles similar to those obtained using X-ray diffraction for both dry and hydrated collagen fibrils[33]. PSHG imaging of collagen in mechanically stretched tendon has shown results consistent with stretching of the collagen molecules[34], indicating that PSHG may be applicable for measuring the molecular strain in stretched collagen fibrils.

Here we perform combined PSHG and AFM measurements of collagen fibrils which have been held under strain on an elastic substrate, thereby allowing simultaneous measurements of the D-band strain, molecular strain, and relative density of collagen molecules. We perform these measurements on fibrils isolated from two different tendon types, one positional tendon with very few mature crosslinks, and one energy storing tendon with a high density of mature crosslinks[35]. We show that in both cases the molecular strain is significantly larger than the D-band strain, with the two fibril types diverging significantly at higher levels of strain. We also observe a transition from an SHG producing triple helical state to a non SHG producing state where the signal is likely lost due to a combination of local loss of triple helix structure and denaturation[36,37]. We further propose that an intermediate state along the transition pathway is a triple helix without its superhelical twist as recently observed by Kreutzberger et al. in the structure of a defense collagen[38]. We additionally show that fibrils with high numbers of mature trivalent crosslinks are more resistant to this strain induced untwisting of the collagen molecular bundle of three chains compared to fibrils with high numbers of immature divalent crosslinks.

## Materials and Methods

**Sample Preparation:** Bovine forelimbs are obtained from a 12 – 18-month-old steer slaughtered for meat at a local abattoir (Oulton Farms, Nova Scotia, Canada), to ensure that the experiment is not affected by differences in nonenzymatic crosslink accumulation with age (e.g. see [25,39,40]) all tendons used here are taken from the same animal. The common digital extensor (positional) and deep digital flexor (energy storing) tendons were dissected from the forelimbs, wrapped in gauze soaked in PBS, and stored at -80° C until use. To isolate fibrils from the tendons a ~2 mm long section of tendon is placed in a dish with ultrapure water. Fiber bundles are removed from the tendon section by pulling with tweezers and then drawn across a rectangular strip of 120 µm thick polydimethylsiloxane (PDMS, Electron Microscopy Sciences) mounted on a glass cover slip (16004-314, VWR), thereby leaving behind individual fibrils which have adhered to the PDMS. The sample is then washed with ultrapure water and left overnight to air dry. To stretch the fibrils the PDMS is manually stretched by 15 – 30 % and secured to the coverslip under strain using tape. The sample is left at room temperature for at least 12 hours before imaging to prevent movement of the sample during data collection. Alternatively, fibrils are adhered to a glass slide then washed with ultrapure water and left overnight to air dry.

**PSHG Imaging and Analysis:** PSHG images are obtained using a custom built laser scanning microscope coupled to an ultrafast laser (FemtoLux 3, EKSPLA) as has been described previously[41,42]. SHG intensity images are obtained at all combinations of 8 linear polarization angles of the excitation laser ($\theta$) and 8 angles of a linear polarizing filter in the detection path ($\phi$) (both angles are measured with respect to the fibril axis). At the end of the imaging process an additional image is obtained at the initial values of $\theta$ and $\phi$ to confirm that there has been no movement or photodamage to the sample during imaging. Under the assumption that the fibrils are composed of a cylindrically symmetric arrangement of uniaxial SHG emitters, the SHG intensity of a fibril can be described by the following equation[43].

$$I \propto N^2|\rho \cos^2(\theta)\cos(\phi) + \sin^2(\theta)\cos(\phi) + \sin(2\theta)\sin(\phi)|^2 \quad (1)$$

Where N is the number of collagen molecules in the fibril and within the focal volume, and $\rho$ is a ratio of molecular second order electric susceptibility tensor elements $\rho = \frac{\chi^{(2)}_{zzz}}{\chi^{(2)}_{zxx}}$ (where z is the fibril axis). Experimental data is fit to equation (1) using a custom MATLAB (Mathworks) program to determine the value of $\rho$ at each pixel. Measurement of the $\rho$ parameter provides information on the molecular structure of the collagen molecule as it is related to the helical pitch angle of the molecule via the following[44,45]

$$\frac{\rho}{2+\rho} \approx \cos^2\psi \quad (2)$$

where $\psi$ is the mean pitch angle of the polypeptide backbone with respect to the molecular axis within the microscope focal volume. This relationship allows us to measure the strain on the polypeptide backbone in stretched collagen fibrils (termed "molecular strain" below) based on the percent change in $\cos\psi$.

PSHG can also be used to determine the relative density between different areas along a fibril. To do this, we sum all the images obtained during the process described above (minus the final error check image) to obtain the total SHG intensity ($I_{sum}$) at each pixel, then by integrating equation (1) over $\theta$ and $\phi$ we obtain:

$$N \propto \sqrt{\frac{I_{sum}}{\rho^2+\frac{2}{3}\rho+\frac{7}{3}}} \quad (3)$$

Comparing the values of N obtained from multiple pixels along a single fibril provides us with the relative density of collagen molecules at each pixel.

**AFM Imaging and Analysis:** Stretched fibrils were imaged using a Bioscope Catalyst (Bruker) atomic force microscope. Images are obtained with a size of 10 × 10 µm in peak force quantitative nanomechanical mapping mode with a pixel size of 10 nm, line scan rate of 0.25 Hz, and peak force setpoint of 5 nN. After each scan the scan window is offset by 9.5 µm along the direction of the fibril and repeated to obtain data along the entire length of the fibril. The images are then flattened using Gwyddion[46] and the individual images are stitched together using FIJI[47,48] for easy comparison with PSHG data. The workflow for obtaining the D-band length is shown in Figure 1. First a line profile of height along the entire length of the fibril is extracted. This data is smoothed by subtracting the moving average within a 150 nm window. The D-band length at each point is determined based on the most intense frequency component of the power spectrum within an 8 μm moving window.

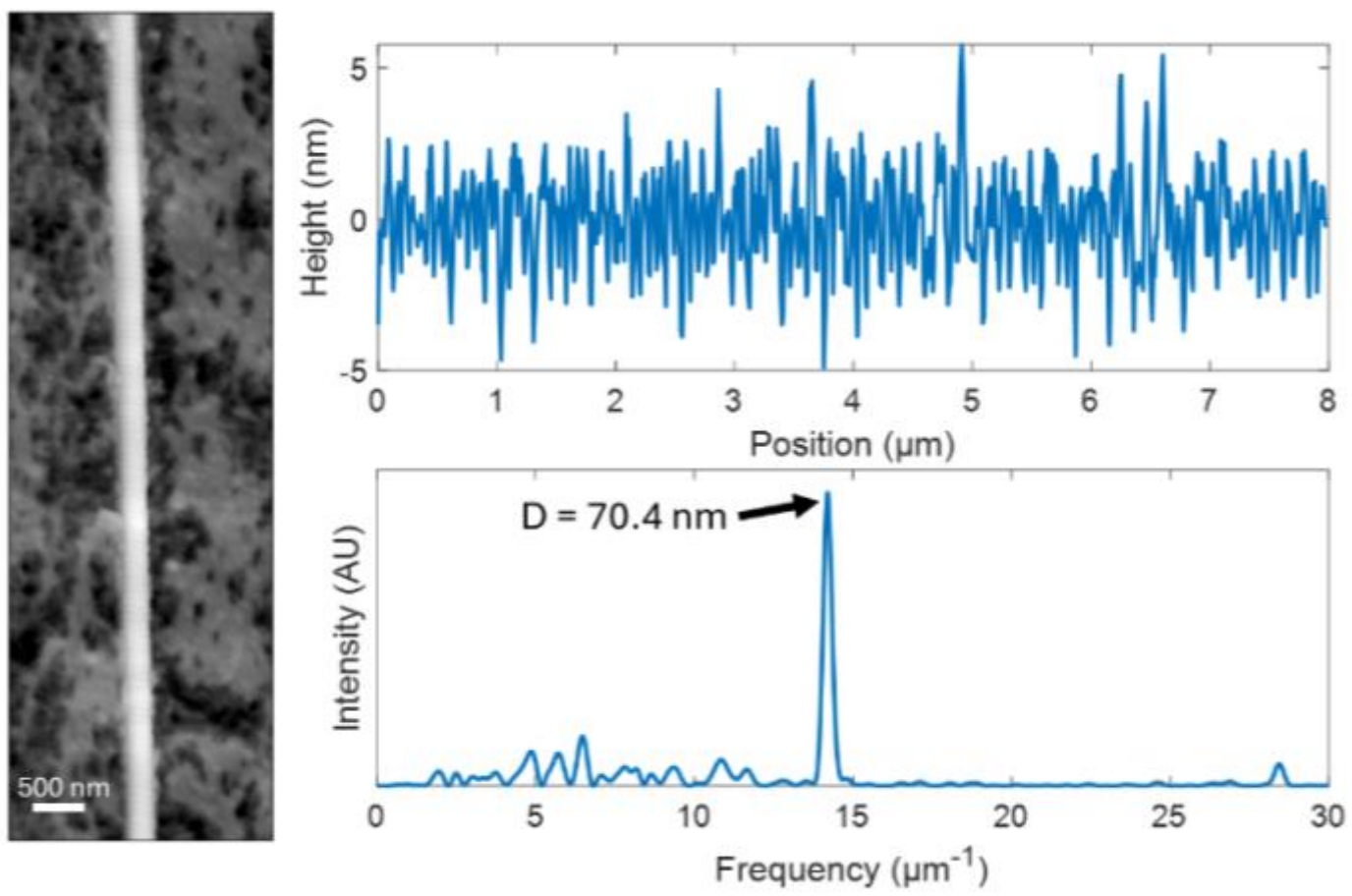

Figure 1: An AFM height image showing an 8 μm long section of a stretched fibril. The plots show the smoothed height profile along the length of the fibril and its corresponding power spectrum. The main peak of the power spectrum corresponds to a D-band spacing of 70.4 nm.

**Theoretical Modeling:** Energy landscape databases have been obtained previously for work described in [49]. Structures from this data set present the equilibrium ensembles, and the data contains their relative potential energies. In addition, a collagen fibril model using the Bos Taurus sequence was generated using ColBuilder[50].

To calculate the second order electric susceptibility tensor, the dominant SHG emitter in proteins is assumed to be the C-N of the peptide bond, with a uniaxial (single element) local frame hyperpolarizability tensor ($\beta_{PB}$) [51–54]. The second order electric susceptibility tensor ($\chi^{(2)}$) of the corresponding triple helix can then be calculated, via a summation of contributions from each peptide bond.

$$\chi^{(2)} = \sum_{\text{All Bonds}} R^{z}(\psi_n)R^{x}(\sigma_n)\beta_{PB} \tag{4}$$

Here $\psi_n$ and $\sigma_n$ are the pitch and azimuthal angles, respectively, of the $n$th peptide bond. $R^{z}$ and $R^{x}$ are standard rotation matrices about the $z$ and $x$ axes, respectively. Therefore, by extracting the angles of all peptide bonds within a given PDB file we are able to calculate the value of $\rho = \frac{\chi^{(2)}_{zzz}}{\chi^{(2)}_{zxx}}$ for that structure. Prior to the computation of rho, the tropocollagen helices were put into a standard orientation as follows. First, the centre of mass was translated to the origin. Then, the largest principal axis was aligned with the z-axis, meaning the helical axis is broadly aligned with the z-axis. As the data from [48] contains the ensemble structures, the computed rho presents the weighted average across the ensemble, where the weight for each structure is the Boltzmann factor at 300 K using the relative potential energy to the global minimum. All scripts and related data are available on github (https://github.com/koroeder/CollPitchAngle).

## Results and Discussion

**Modeling Collagen PSHG:** To confirm the ability of PSHG to estimate molecular strain in stretched collagen fibrils we begin by performing theoretical modeling to determine the ρ values from energy landscape models of a collagen model peptide under strain (See structures in Figure 2) previously published in[49]. These simulations are important as equation (2) which relates the measured $\rho$ to the mean pitch angle of the molecule is based on the assumption that the distribution of pitch angles within the molecule is narrow. Previous theoretical work has shown that broad Gaussian distributions of pitch angles can lead to inaccurate estimates of the mean angle[55]. Therefore, simulations using a realistic pitch angle distribution are needed to demonstrate the accuracy of the experimental measurements presented below. The calculated ρ value for the unstrained collagen model peptide was 1.324 ± 0.009, which is slightly smaller than the value of 1.43 previously measured for hydrated collagen fibrils[33], however, given the high content of proline and hydroxyproline in our model compared to biological collagen a higher helical pitch angle (and therefore a lower ρ value) is expected[56]. Using a more realistic model of the collagen fibril (see Supplement and Figure S1), we obtain a ρ value of 1.98, similar to the previously measured value of 2.00 in dried collagen fibrils[33].

For the collagen model peptide under strain, as shown in a cartoon representation in Fig. 2a we find little change in the estimated pitch angle over the first 100 pN of applied tension (see Figure. 2b). This is expected as for low strain the main deformation mechanism of the collagen molecule is unwinding of the triple helix[57,58], which narrows the width of the pitch angle distribution without changing the mean value. At 250 pN the pitch angle distribution becomes significantly narrower, and the mean pitch angle, as well as the estimated pitch angle begin to decrease as the backbone of the helices begins to stretch. At this point the difference between the true mean pitch angle of the molecule and the pitch angle estimated from ρ becomes much smaller, indicating that PSHG can be used to accurately measure the strain on the polypeptide backbone of individual single helices during stretching of the collagen molecule.

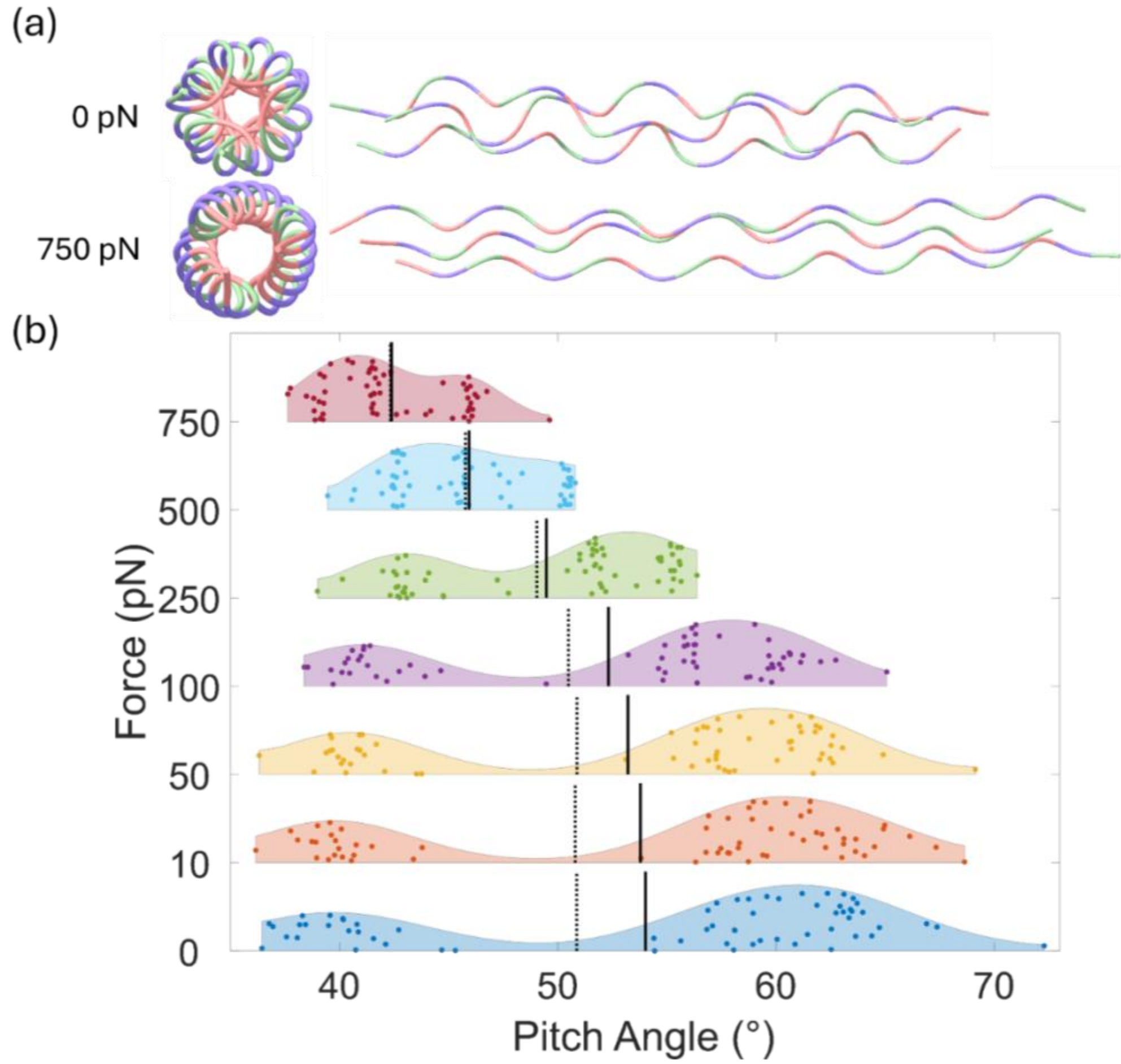

Figure. 2: Effect of molecular stretching on PSHG measurements. Images showing cartoon representations of a collagen molecule under 0 and 750 pN of applied tension (a). Glycine is represented in pink, proline in green, and hydroxyproline in purple. Chart showing the distribution of peptide bond pitch angles for all the simulated molecules (b). The vertical solid lines show the mean pitch angles, while the dotted lines show the angle calculated from the simulated ρ value of each structure.

**PSHG Imaging of Stretched Fibrils:** To investigate the effect of stretching on collagen molecular structure we perform combined PSHG and AFM imaging of collagen fibrils stretched on an elastic substrate. To investigate the effect of crosslinks on collagen mechanical response we image three random fibrils isolated from a positional tendon, since these are known to contain mainly divalent immature crosslinks, and three from an energy storing tendon, which are known to contain a high density of mature trivalent crosslinks[35]. For imaging, we select collagen fibrils that are roughly aligned with the long axis of the PDMS strip and close to its edges. These fibrils show a gradient of SHG intensity along their length (Figure 3a, b) which correlates with a gradient in ρ (Figure 3c) and a gradient in D-band length (Figure 3d), indicating that there is a gradient of strain along the length of the fibril. The SHG intensity, ρ and D-band lengths of measured fibrils are summarized in Table 1.

| | Positional | | | Energy Storing | | |
|---|---|---|---|---|---|---|
| | Intensity Range (normalized) | ρ Range | D-band Range (nm) | Intensity Range (normalized) | ρ Range | D-band Range (nm) |
| Fibril 1 | 0.25-1 | 2.05-4.73 | 65.24-71.54 | 0.07-0.30 | 2.15-2.64 | 63.49-66.51 |
| Fibril 2 | 0.75-1 | 3.31-6.71 | 71.19-73.90 | 0.07-0.57 | 2.15-4.30 | 63.38-71.03 |
| Fibril 3 | 0.59-0.86 | 6.28-10.94 | 73.88-75.65 | 0.44-1 | 2.99-7.02 | 69.24-77.15 |

We observe up to a four-fold increase in SHG intensity for positional fibrils and greater than a tenfold increase for energy storing fibrils. The maximum D-band lengths are 75.65 nm for positional and 77.15 nm for energy storing fibrils. This is typical of other studies that have used the same approach to stretch single fibrils[22,23]. Considering that the D-band length of a tendon fibril at rest is 66 nm[59], this shows that the SHG intensity of a single collagen fibril can be strongly modulated over a wide range of strains. Even though the SHG intensity increases observed here are modest compared to strain induced enhancement observed in 2D layered materials[60,61], collagen fibrils could serve as inspiration to design stretchable polymer based SHG emitters[62].

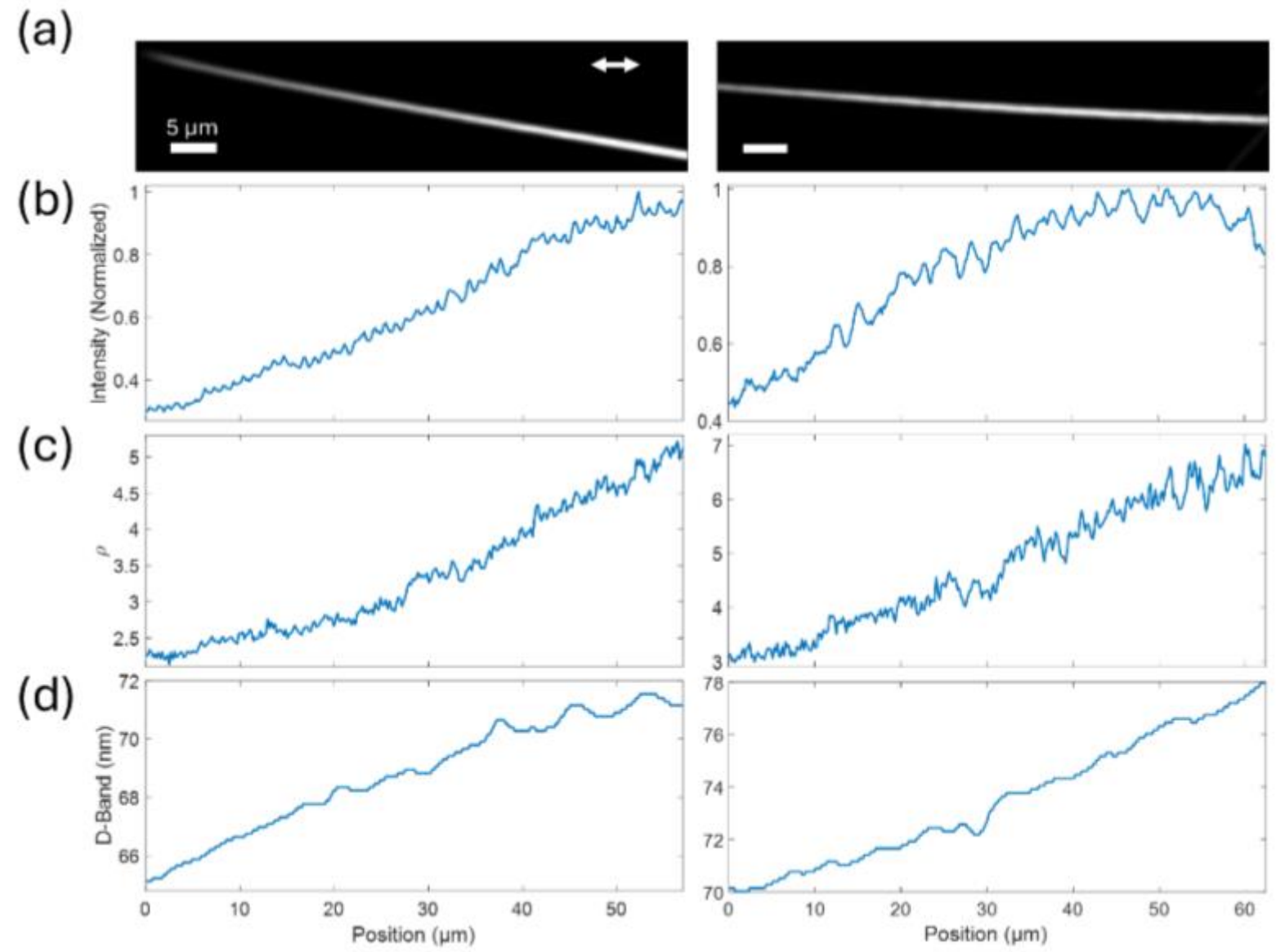

Figure 3: Examples of stretched collagen fibrils showing a low strain region near the end of a positional fibril (left) and a high strain region in the middle of an energy storing fibril (right). (a) SHG intensity images obtained by adding all 64 polarization images, the white double-headed arrow indicates the stretching direction. (b) SHG intensity normalized to the maximum value as a function of position along the fibrils in (a). (c) Anisotropy parameter ρ fitted at each pixel along the fibrils in (a) using equation (1). (d) D-band extracted from the height profile along the fibrils corresponding to the SHG images in (a).

**Positional and Energy Storing fibrils show different deformation pathways:** Next, we want to explore the deformation pathways of positional and energy storing fibrils. To do so, we need to first define the rest state of each fibril-type in order to compute molecular strain and D-band strain. We used PSHG to image 3 positional and 3 energy storing fibrils deposited on a glass slide and obtained a mean $\rho_0 = 1.9 \pm 0.1$ and $2.1 \pm 0.1$ (where $\pm$ is standard deviation), respectively. The D-band value at rest was set to 66 nm which is a well-accepted value for single tendon fibrils. We did not attempt to estimate this value by AFM imaging as the measured value is very sensitive to tip shape and has a standard deviation of a few nanometers when comparing multiple fibrils[59]. One reason for this variability is that fibrils extracted from tendons have some variable degree of internal strain as proposed by Dittmore et al[63]. This may also explain why the stretched fibrils seem to have accumulated 3-4 % molecular strain at 0 % D-band strain (Figure 4a, Figure S2), or it is simply an artefact of our conservative choice of a reference D-band length. Note that varying the D-band value at rest between 65 and 67 nm does not change the results presented in Figure 4 qualitatively.

For both positional and energy storing fibrils, we observe that the molecular strain increases faster than the D-band strain (Figure 4a). The ratio is 1.3 for the two fibril types below 5 % D-band strain and it even increases to 2.2 for the positional fibrils above 5 % D-band strain (Figure 4a). Note that at around 15 % D-band strain the molecular strain appears to plateau at about 20 % for energy storing fibrils and 30 % for positional fibrils. These values are large but below the theoretical limit for a fully straightened triple helix which should be about 33 % assuming a rise per residue of 0.286 nm[64] for the collagen triple helix and a rise per residue of 0.38 nm for a fully straightened polypeptide[65]. In comparison, molecular strain measurements performed at the tendon level using X-ray scattering always show that the molecular strain is lower than the D-band strain but is limited to about 2.5 % molecular strain[24,28].This is because X-ray scattering requires that all the triple helices deform somewhat uniformly in order to measure a change in the rise per residue. In contrast PSHG captures scattering from all molecules in the focal volume even if they carry different molecular strain and are significantly distorted as long as the local structure remains noncentrosymmetric.

Considering that we are directly stretching the fibrils on a substrate, it seems appropriate to equate D-band strain with fibril strain. In that case, energy storing fibrils appear to support almost affine deformation up to about 15 % fibril strain with a ratio of molecular strain to D-band strain of 1.3 (Figure 4a). This implies that mature trivalent crosslinks allow strain to distribute uniformly within the fibrils while restricting molecular strain to only 20 % compared to 30 % for immature divalent crosslinks (see the maximum values of molecular strain in Fig. 4a). In comparison positional fibrils support affine-like deformation up to 5 % fibril strain (Figure 4a) and past that point they start rapidly accumulating molecular strain which is likely due to strain concentration as previously observed in molecular dynamics simulations[14]. More specifically, simulations predict that above 7 % fibril strain (phase II in [14]) molecules should start sliding past each other, at least in the case of a fibril with immature divalent crosslinks. Given that triple helices are likely at least partially uncoiled before the molecular backbone is strained (see Modeling Collagen PSHG above), it is interesting that we are able to observe backbone strains of up to 30 %. This indicates that the collagen remains relatively stable even after uncoiling of the molecules, in agreement with previous measurements of collagen in mechanically overloaded tendons[66].

To try to capture evidence of molecular sliding experimentally, we compute the relative density of emitters using equation (3) as a function of D-band strain (Figure 4b, Figure S2). At D-band strain below 5 %, the density of emitter increases in agreement with molecular dynamics simulation (phase I in [14]) that predict a straightening of pre-existing molecular kinks, which will better align the molecules to the fibril axis resulting in more efficient SHG. Above 5 % D-band strain, positional fibrils show an abrupt decrease in emitter density that can be interpreted in two different ways that are not mutually exclusive. One possibility supported by simulations and X-ray scattering data on stretched tendon[25] is that molecules start sliding past each other which decreases the number of emitters in the focal volume, or parts of the molecules stop producing SHG altogether due likely to bond breaking. The latter has only been observed when collagen samples are heated above their denaturation temperature[67]. Independently of the mechanisms at play, they appear to be significantly restricted and delayed in energy storing fibrils (Figure 4b).

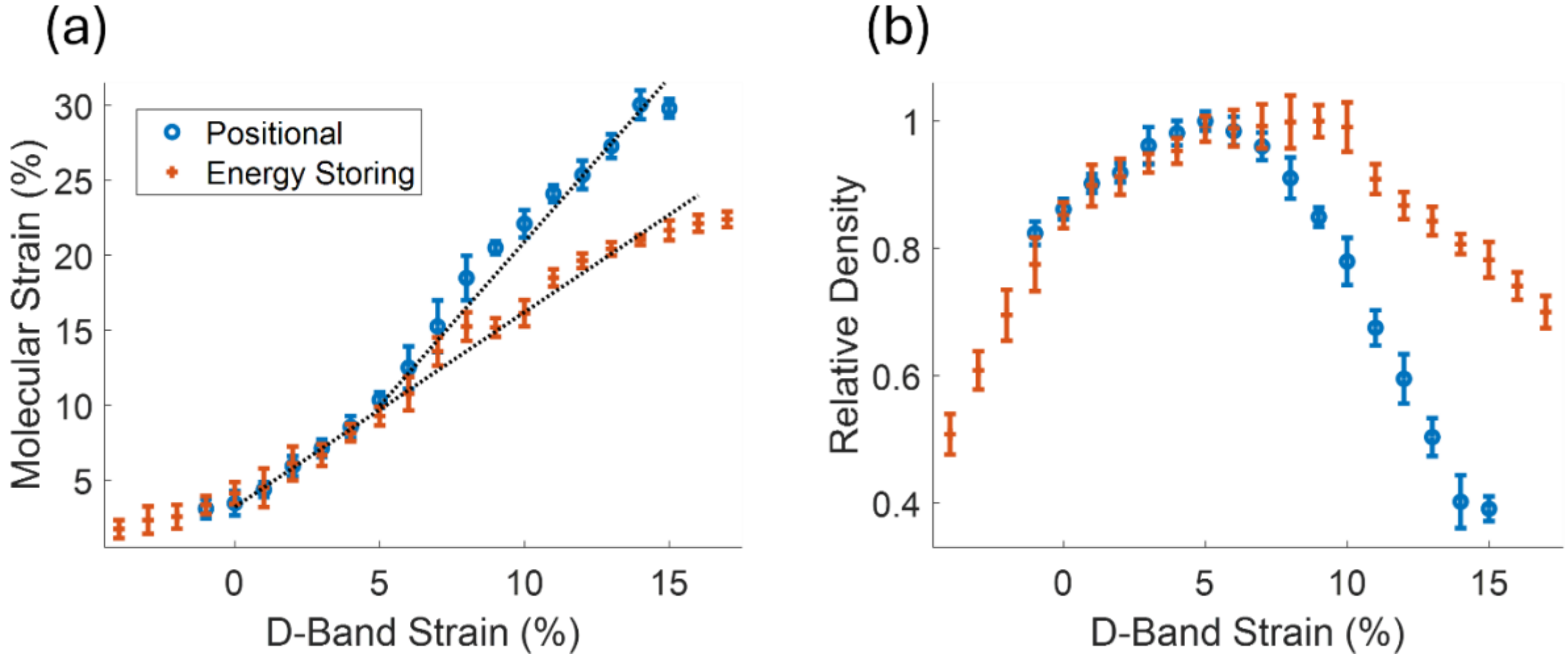

Figure 4: Deformation pathways of positional (open symbols) and energy storing (crosses) fibrils. (a) Molecular strain obtained from equation (2) as a function of D-band strain for n =3 fibrils. At positive D-band strain the molecular strain exceeds the D-band strain for all fibrils. The ratio is 1.3 for the energy storing fibrils (dotted line) and for the positional fibrils up to about 5 % D-band strain. The ratio is 2.2 for the positional fibrils above 5 % D-band strain. (b) Relative density of emitters obtained from equation (3) as a function of D-band strain for n =3 fibrils. Both fibril types show an increase in emitters density followed by a plateau and a decrease. The decrease is sharper and occurs at lower D-band strain in the positional fibrils compared to the energy storing fibrils.

**PSHG uncovers a structural transition in stretched collagen fibrils:** To try to decouple molecular sliding from molecular unfolding we plot the density of emitters N as a function of molecular strain for each fibril type (Figure 5). Above 10 % molecular strain both curves are remarkably similar and appear to follow a sigmoidal curve (Figure 5). If this is due to the unfolding of the collagen triple helices under strain, then it can be modelled by a two-state system where the free energy barrier decreases with strain. Applying Boltzmann statistics, the density of native-state emitters should be described by:

$$N = \frac{1}{1+e^{-\frac{(\Delta G_0 - k\varepsilon)}{k_B T}}} \tag{5}$$

This simple model fits both curves well (Figure 5) and highlights subtle differences between the two fibril types. The energy barrier at rest $\Delta G_0$ is higher for energy storing fibrils at 10.2 ± 1.5 $k_B T$ compared to positional fibrils at 6.3 ± 0.6 $k_B T$. However, the rate of energy barrier decrease per percent strain $k$ is also larger for energy storing fibrils at 0.42 ± 0.07 $k_B T$/% compared to positional fibrils at 0.23 ± 0.02 $k_B T$/%. Altogether this means that the transition is both delayed and more abrupt or cooperative in energy storing fibrils compared to positional fibrils. The delay can be attributed to trivalent crosslinks restricting collagen triple helices stretching more than divalent crosslinks, whereas the cooperativity is consistent with trivalent crosslinks yielding a more uniform strain distribution within the fibril compared to divalent crosslinks (Figure 4a).

There are two limitations to these findings. First instead of considering the density of emitters we should be considering their number and correct for any change in volume due to stretching. We attempted to measure the height of the fibril as a function of position to estimate the change in cross-sectional area, but the observed trend was dominated by the behaviour of the substrate indicating that either the cross-section of the fibril did not change or most likely that the change was too small to be robustly detected. We then attempted to measure a decrease in the D-band height amplitude as a function of D-band length as previously observed by Wenger and Mesquida[68] by computing the Fourier amplitude of the D-band peak but observed no significant trend (see Fig S3 in the Supplement). Without a clear measurement, we did not attempt any further corrections. Second the density of emitters and the molecular strain are two quantities extracted from the same PSHG measurement so ideally this transition should be confirmed with another experimental technique to provide additional information on the nature of the transition.

The extracted free-energy barriers of 6–10 $k_B T$are consistent with the cooperative destabilization of a localized triple helical segment rather than complete denaturation of the collagen molecule. Global thermal unfolding involves the cooperative disruption of stabilizing interactions over the entire ~300 nm triple helix and occurs on a much larger energetic scale[69]. We therefore interpret the observed transition as a localized, strain-induced conformational perturbation of the triple helix capable of decreasing SHG emission. We propose that the collagen triple helices composed of three polyproline II helices with a supertwist locally transition to a straight bundle conformation like the one recently observed by Kreutzberger et al. in the defense collagen C1q[38]. As these segments become untwisted and taught, it localizes stress to weak bonds that break, as proposed in molecular dynamics simulations[20], leading to the denaturation of large fragments of the collagen molecules. This in turn explains the significant loss of emitter density at large molecular strains (Figure 5). The untwisting of the superhelical twist fits with the low energy barrier observed as well as the transition onset. Considering that the rise per residue in a straight polyproline II helix is 0.31 nm[70] then removal of the triple helix supertwist should start above 8.4 % which matches well with our observations for both fibril types (Figure 5). The fact that this transition may occur only at specific sites along the sequence is in agreement with recent observations that local bending[37] and thermal denaturation[36] properties are also sequence dependent in collagen molecules.

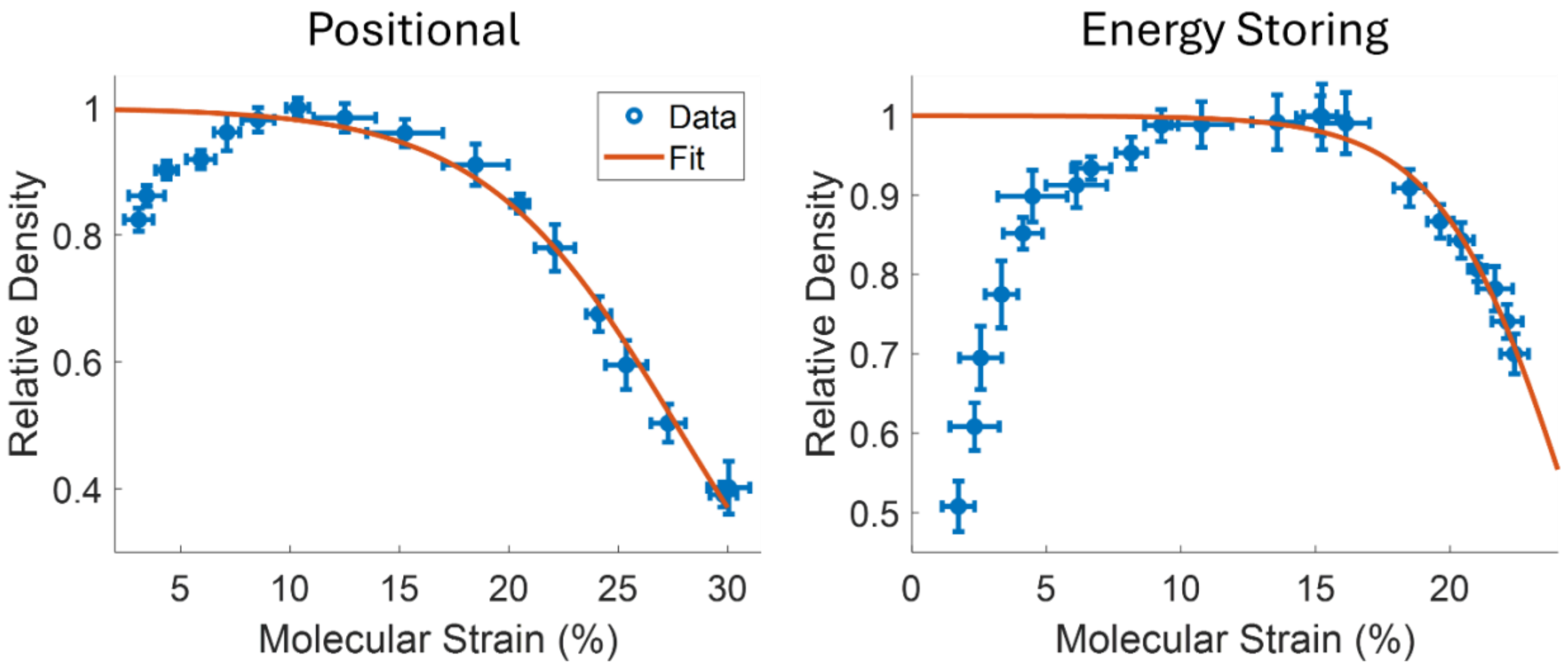

Figure 5: Density of emitters as a function of molecular strain for positional and energy storing fibrils. In both cases the data is fit above 10 % molecular strain using equation (5).

## Conclusions

Combining single collagen fibril stretching on an elastic substrate along with co-registered PSHG and AFM measurements, we have revealed that collagen fibrils are strain tunable SHG emitters up to 15 % strain. By analyzing the polarization dependence of the SHG intensity we were able to show significant change in the ρ parameter. This change in ρ was leveraged to measure molecular strain levels within positional fibrils with immature divalent crosslinks and energy storing fibrils with mature trivalent crosslinks. We showed that trivalent crosslinks restrict molecular stretching and yield a more homogenous molecular strain distribution compared to divalent crosslinks. Finally, we uncover a structural transition happening above 10 % molecular strain for both fibril types that we hypothesize to involve in part the local untwisting of the collagen triple helix supertwist. Our results open the road for the development of strain tunable SHG emitting polymers. They also shed new light on the complex deformation pathway of collagen molecules within fibrils and offer a direct bridge with fibril scale molecular dynamics simulations.

**Funding:** Natural Sciences and Engineering Research Council of Canada (RGPIN-2018- 03781 for L.K. and RGPIN-2018-05444 for D.T.); Canada Foundation for Innovation (John R. Evans Leaders Fund #37749 for D.T.); Research Nova Scotia (Doctoral Scotia Scholars Award for M.H. and 1868 for D.T.); Nova Scotia (Nova Scotia Graduate Scholarship for M.H.); Canada's Research Support Fund; Saint Mary's University.

Supplementary file

# Polarization controlled second harmonic generation imaging of stretched collagen fibrils reveals collagen's deformation pathway in situ

MacAulay Harvey, Konstantin Röder, Richard Cisek, Danielle Tokarz*, Laurent Kreplak*

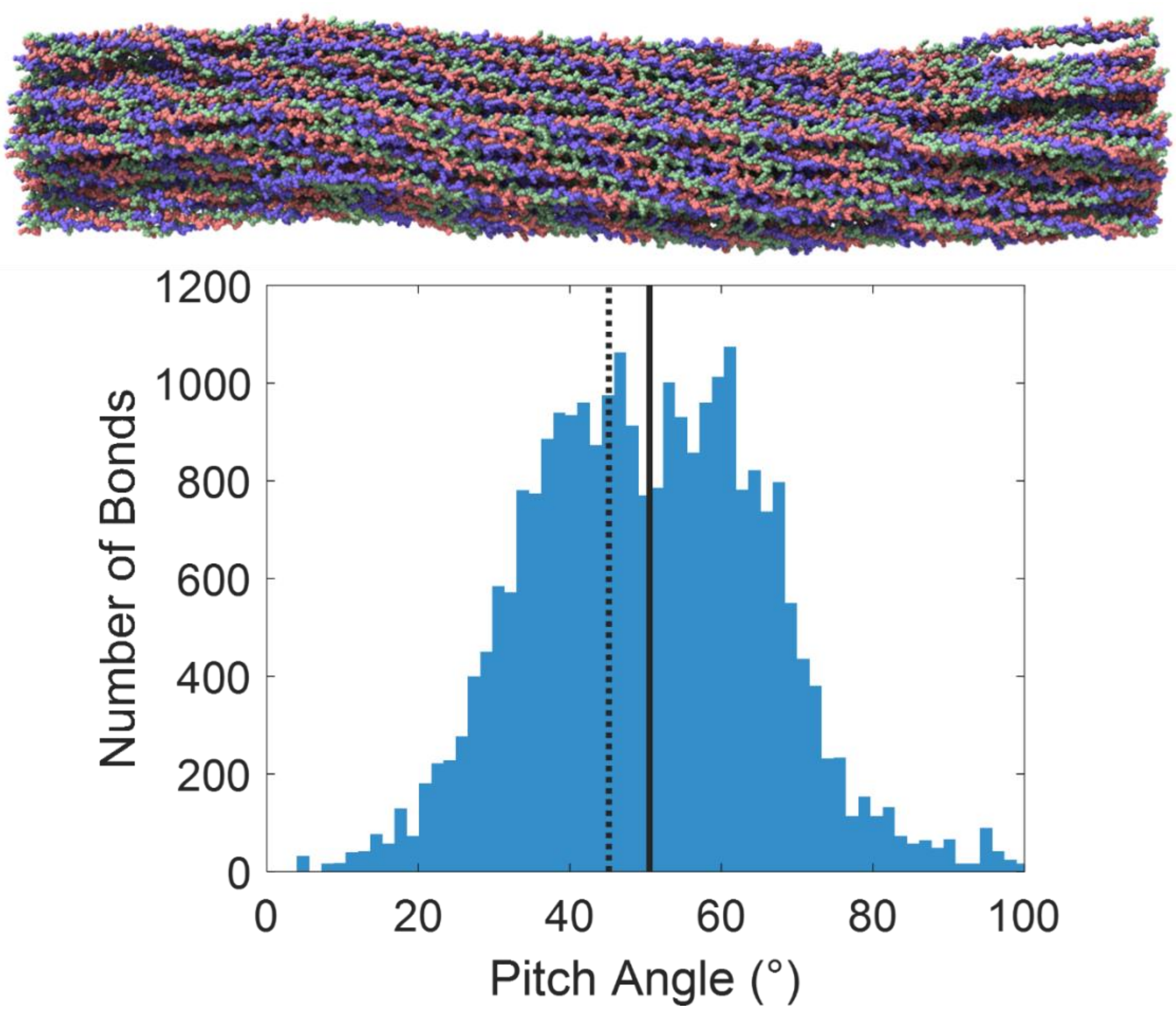

Figure S1: Collagen fibril fragment generated with ColBuider using the bos taurus sequences for the chains. The three chains of each triple helix are represented in different colours. The PDB file generated by ColBuilder[1] was used to generate a histogram of all pitch angles ($\psi$) for the fibril fragment comprising a total of 25979 peptide bonds. The dotted black line represents the $\psi$ value calculated using equation (2) and ρ = 1.98, estimated using the PDB file and equation (4), and the solid black line is the true mean values of $\psi$. Note that ρ does not exactly measure the average angle $\psi$ of the structure as it is also sensitive to the width of the angle distribution.

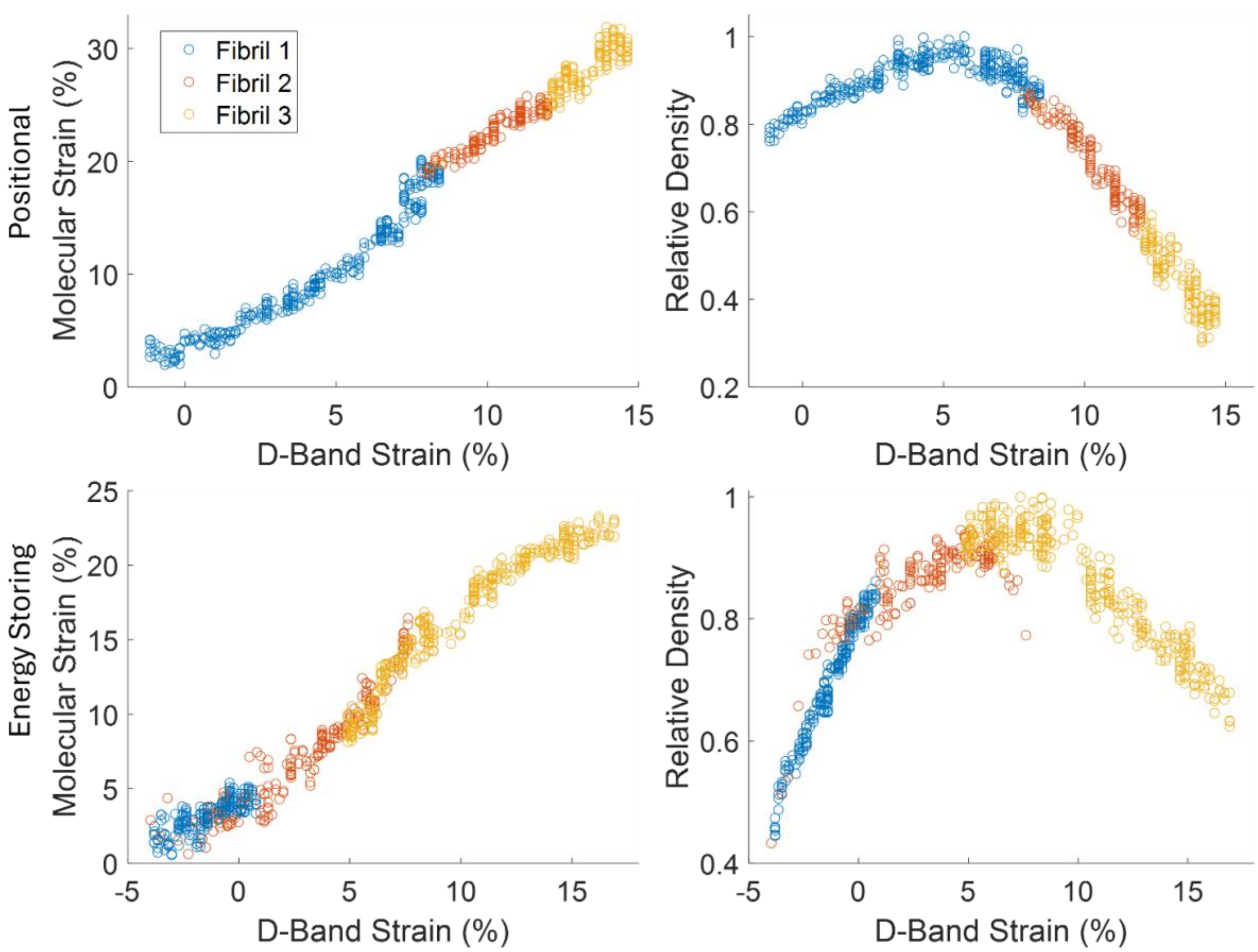

Figure S2: Molecular strain and relative density as a function of D-band strain for positional and energy storing fibrils before binning (see Figure 4). Here the data is broken down on a fibril per fibril basis. The numbering of fibrils is the same as Table 1.

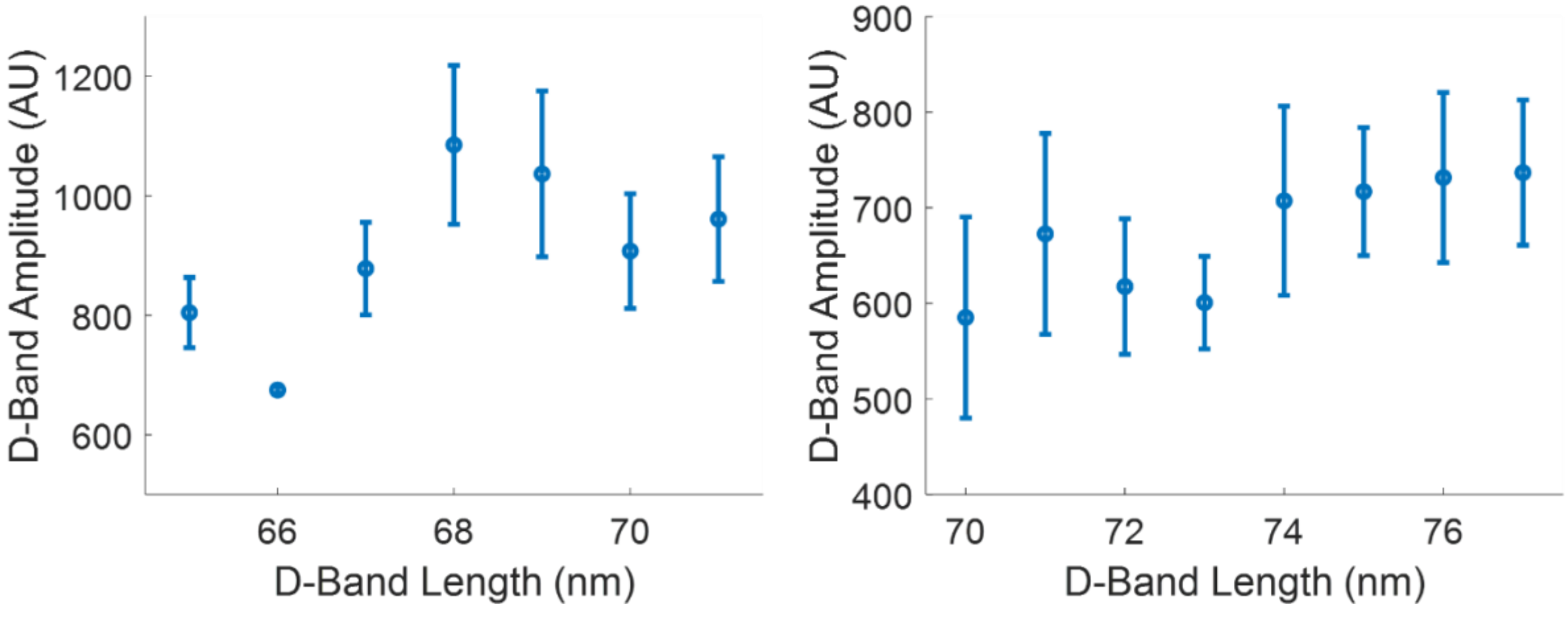

Figure S3: Amplitude of the power spectrum (mean ± standard deviation) as a function of D-band length for the two fibrils shown in Fig. 3 of the main text. The data is binned in increments of 1 nm in D-band length.